\documentclass[useAMS,usenatbib,fleqn,twocolumn]{aastex631}
\usepackage{multirow}
\usepackage{graphicx}
\usepackage{xcolor,soul}
\usepackage{amssymb,amsmath}
\usepackage{aas_macros}
\usepackage{tabularx} 
\usepackage[T1]{fontenc} 
\begin{document}

\shorttitle{Super-Eddington Spin Evolution}
\title{Recipes for Jet Feedback and Spin Evolution of Black Holes with Strongly-Magnetized Super-Eddington Accretion Disks}

\shortauthors{Ricarte, Narayan, \& Curd}
\correspondingauthor{Angelo Ricarte}
\email{angelo.ricarte@cfa.harvard.edu}

\author[0000-0001-5287-0452]{Angelo Ricarte}
\affiliation{Black Hole Initiative at Harvard University, 20 Garden Street, Cambridge, MA 02138, USA}
\affiliation{Center for Astrophysics | Harvard \& Smithsonian, 60 Garden Street, Cambridge, MA 02138, USA}

\author[0000-0002-1919-2730]{Ramesh Narayan}
\affiliation{Black Hole Initiative at Harvard University, 20 Garden Street, Cambridge, MA 02138, USA}
\affiliation{Center for Astrophysics | Harvard \& Smithsonian, 60 Garden Street, Cambridge, MA 02138, USA}

\author[0000-0002-8650-0879]{Brandon Curd}
\affiliation{Department of Physics $\&$ Astronomy, The University of Texas at San Antonio, One UTSA Circle, San Antonio, TX 78249, USA}

\date{\today}

\begin{abstract}
A spinning black hole accreting from a disk of strongly magnetized plasma via a magnetically arrested disk is known to produce an efficient electromagnetic jet powered by the black hole's spin energy.  We present general relativistic radiative magnetohydrodynamic simulations of magnetically arrested systems covering a range of sub- to super-Eddington accretion rates. Using the numerical results from these simulations, we develop formulae to describe the magnetization, jet efficiency, and spin evolution of an accreting black hole as a function of its spin and accretion rate.  A black hole with near-Eddington accretion experiences a mild degree of spin-down because of angular momentum loss through the jet, leading to an equilibrium spin of 0.8 rather than 1.0 at the Eddington limit. As the accretion rate increases above Eddington, the spin-down effect becomes progressively stronger, ultimately converging on previous predictions based on non-radiative simulations.  In particular, spin evolution drives highly super-Eddington systems toward a black hole spin near zero.  The formulae developed in this letter may be applied to galaxy and cosmological scale simulations that include black holes.  If magnetically arrested disk accretion is common among supermassive black holes, the present results have broad implications for active galactic nucleus feedback and cosmological spin evolution.
\end{abstract}


\section{Introduction}
\label{sec:introduction}

Astrophysical black holes (BHs) accreting from disks of plasma are known to launch relativistic jets and outflows \citep{Fabian2012,Heckman&Best2014}.  Such energy injection from supermassive BHs (SMBHs) at the centers of galaxies, a process referred to as active galactic nucleus (AGN) feedback, is believed to be essential for stopping runaway gas cooling and star formation in massive galaxies and dark matter halos \citep{DiMatteo+2005,Springel+2005,Croton+2006,Sijacki+2007,Kormendy&Ho2013,Harrison+2017}.  In this paradigm, accretion and feedback processes are critical for a complete picture of SMBH growth and galaxy co-evolution. However, the details remain poorly understood.

For magnetized accretion disks, an electromagnetic analogue of the \citet{Penrose1969} process known as the \citet{Blandford&Znajek1977} (BZ) mechanism provides the most widely accepted model for jet launching.  The power of a jet launched by the BZ mechanism scales approximately proportional to both the square of the BH spin and the square of the magnetic flux threading the horizon.  In systems with high enough spin and with maximal magnetic field strength, corresponding to a so-called magnetically arrested disk (MAD) \citep{Bisnovatyi-Kogan&Ruzmaikin1974,Igumenshchev+2003,Narayan+2003}, more jet power can be launched than the entire rest mass energy of the material flowing into the BH \citep{Tchekhovskoy+2011}.  The extra energy is supplied by the spin kinetic energy of the BH, which thereby may cause the BH to spin down with time.  In this way, jets that travel through dark matter halos for hundreds of kiloparsecs are ultimately linked to the evolution of BH spin and the transport of magnetic fields on event horizon scales.  

Since the BZ mechanism powers a jet by extracting BH spin energy, if the process continues long enough a BH could continuously spin down and equilibrate near a spin value $a_* \approx 0$. This has been explicitly demonstrated via general relativistic magnetohydrodynamic (GRMHD) simulations of radiatively inefficient, geometrically thick, MAD models \citep{McKinney+2012,Tchekhovskoy+2012,Narayan+2022,Lowell+2023}.  Several recent publications have begun to study the implications of this spin-down effect for BH populations over cosmic time.  The systems simulated so far largely belong to the regime of advection-dominated accretion \citep{Narayan-Yi1994,Narayan-Yi1995}, or hot accretion \citep{Yuan&Narayan2014}, which corresponds to highly sub-Eddington accretion. Spin-down is relatively slow for such low Eddington-ratio systems simply because the mass accretion rate is very small; nevertheless, continuous jet feedback from such BHs is implicated for maintaining low star formation for Gyrs in some galaxies \citep[e.g.,][]{Hlavacek-Larrondo+2015}, which can lead to cosmologically significant BH spin evolution \citep{Narayan+2022}. 

Super-Eddington accretion disks are geometrically thick and advection-dominated, just like low-Eddington ratio hot accretion flows, and can also reach the MAD state \citep{McKinney+2015,Narayan+2017,Curd+2019}.  Such systems can produce extremely powerful jets \citep[e.g.,][]{Curd+2019}, and because of the very large accretion rate their BHs could spin-down very rapidly.  \citet{Lowell+2023} developed a physical semi-analytic model for this spin-down phenomenon. Using this model, \citet{Jacquemin-Ide+2023} predicted rapidly decreasing collapsar BH spins to $a_* \lesssim 0.2$ near birth.  

Self-consistent BH spin evolution is now being implemented in some galaxy and cosmological-scale simulations, which may then be used to model radiative efficiency and jet power \citep{Dubois+2014,Fiacconi+2018,Bustamante&Springel2019,Beckmann+2019,Dubois+2021,Talbot+2021,Massonneau+2023b,Dong-Paez+2023}.  Although galaxy-scale simulations cannot possibly resolve accretion disk scales, such an approach still represents a substantial improvement over most contemporary work to link SMBH spin evolution to the angular momentum of resolved gas on scales of parsecs.  \citet{Dubois+2021} and \citet{Massonneau+2023b} implement spin-down during periods of thick disk accretion, employing fitting functions for the magnetic flux as a function of spin from GRMHD simulations.  Again assuming the same results that have been demonstrated for very low Eddington ratio disks also hold for super-Eddington disks, \citet{Massonneau+2023b} consider super-Eddington growth in high-redshift galaxies.  While spin-down is noticeable in this simulation, it is counteracted by periods of thin disk accretion.

All such calculations require some a priori knowledge or assumptions about the magnetic field strength.  For magnetized geometrically thick disks in the low-Eddington rate limit, the MAD model offers one well-studied solution.  In contrast to the weak-field ``Standard and Normal Evolution'' (SANE) model \citep{Narayan+2012,Sadowski+2013}, a MAD system is characterized by such strong magnetic fields that magnetic pressure and tension is comparable to the gas pressure near the horizon \citep{Bisnovatyi-Kogan&Ruzmaikin1974,Igumenshchev+2003,Narayan+2003}.  MAD models are characterized by a dimensionless magnetic flux parameter $\phi$ (defined in \autoref{eq::madparam}) saturating at a spin-dependent maximum value \citep{Tchekhovskoy+2012,Narayan+2022}, as well as ``flux eruption events'' that occur when the BH expels magnetic flux \citep[e.g.,][]{Tchekhovskoy+2011,Dexter+2020,Ripperda+2022,Chatterjee&Narayan2022}. The saturated fields that characterize the MAD state lead to highly efficient jets powered by the BZ mechanism.

Spatially resolved and polarimetric observations of the nearby low-luminosity AGN, M87* and Sgr A*, currently favor MAD models over their SANE counterparts \citep{EHT8,EHT_SgrA_V,Wielgus+2022}, suggesting that the saturated values of $\phi$ characteristic of MAD models are easily achieved in low Eddington-ratio geometrically thick hot accretion disks. However, it remains to be confirmed that the same saturation values found for hot accretion flows at low Eddington ratios also hold for super-Eddington accretion flows where radiation plays an important role. It is also unknown whether the BZ mechanism operates efficiently in such systems and how efficiently BH spin-down proceeds. We explore these questions here.

In this letter, we introduce and analyze a suite of super-Eddington general relativistic radiative magnetohydrodynamic (GRRMHD) simulations in the MAD regime to explicitly calculate the magnetization $\phi$, jet power $P_\mathrm{jet}$, and spinup parameter $s$ (defined in equation \autoref{eqn:s_def}), as a function of the dimensionless BH spin parameter $a_*$ and the Eddington ratio $f_\mathrm{Edd}$ (defined in \autoref{eq:fEdd}) of the accretion flow.  As we shall show, highly super-Eddington accretion disks ($f_\mathrm{Edd} \gg 1$) behave similarly to their very low Eddington-ratio ($f_\mathrm{Edd} \ll 1$) counterparts.  However, we find reduced magnetization and spin-down for Eddington ratios $f_\mathrm{Edd} \lesssim 10$.  Based on this behavior, we devise fitting functions for jet power and spin evolution that can be adapted into cosmological and galaxy-scale simulations.

\section{GRRMHD Simulations}
\label{sec:methodology}

Radiation plays a critical role in the dynamics of BH accretion disks for Eddington-ratios $f_{\rm Edd} \gtrsim 0.01$. In these systems, radiative cooling acts to thin the disk at lower Eddington ratios, while radiative pressure puffs up the disk vertically as the mass accretion rate approaches or exceeds Eddington \citep{Abramowicz+1988}. In super-Eddington systems, winds and jets driven purely by radiation can also occur \citep{Sadowski+2015c,Coughlin+2020}

The numerical treatment of radiation in BH accretion problems is quite difficult as the algorithm must treat both optically thin and thick regions in a curved spacetime. \citet{Ohsuga+2005,Ohsuga+2011} pioneered global, non-relativistic, radiation hydrodynamics (RHD) simulations of super-Eddington accretion disks using flux-limited diffusion. Following this work, radiation was first included in the fully general relativistic radiation magnetohydrodynamics (GRRMHD) code, {\sc koral}, by \citet{Sadowski+2013,Sadowski+2014} using the M1 closure scheme and a semi-implicit method to handle the radiation terms. Since then, the M1 closure scheme has been applied in other GRRMHD codes \citep{McKinney+2014,Takahashi+2016,Asahina-Ohsuga2022,Utsumi+2022} as well as a GPU accelerated GRRMHD code \citep{Liska+2023}. Alternative methods of treating radiation in GRRMHD include directly solving the radiative transfer equations to obtain the Eddington tensor \citep{Asahina-Ohsuga2022}, Monte Carlo methods \citep{Ryan+2015}, or using a discretized radiation tensor \citep{White+2023}. The M1 closure scheme allows limited treatment of anisotropic radiation fields. It is superior to the Eddington approximation in optically thin regions, and is well suited for global GRRMHD simulations of super-Eddington disks. However, for complicated radiation fields, it cannot match methods based on the full Eddington tensor.

\citet{Utsumi+2022} explored the role of BH spin in super-Eddington accretion by running a suite of 2D GRRMHD simulations for different spin values. They considered the SANE regime of accretion for which 2D simulations are sufficient. The MAD accretion regime, however, requires 3D simulations and this is the focus of our work.  We present a suite of 38 3D numerical simulations of near-Eddington to super-Eddington MAD simulations carried out with the GRRMHD code, {\sc koral} \citep{Sadowski+2013,Sadowski+2014,Sadowski+2015a,Sadowski+2015b}. We include 2 BH masses, $M = 10,\,10^4 M_\odot$, 6 BH spin values, $a_*=$ -0.9, -0.68, 0, 0.68, 0.9, and 0.97 (where a minus sign denotes retrograde accretion), and a range of Eddington ratios, $0.4 \lesssim f_\mathrm{Edd} \lesssim 40$.  Since prolonged super-Eddington accretion is often invoked for the growth of BH seeds in the early universe, as we will later explore in \autoref{sec:conclusion}, these two masses are loosely motivated by exploring both ``light'' and ``heavy'' seeding scenarios \citep[see e.g.,][for a review]{Natarajan2014}. We define $f_{\rm{Edd}}$ as follows,
\begin{equation}
f_{\rm{Edd}} = \dot{M}/\dot{M}_{\rm{Edd}},
\label{eq:fEdd}
\end{equation}
where $\dot{M}$ is the mass accretion rate through the BH horizon (\autoref{eq:mdotin}) and $\dot{M}_{\rm{Edd}}$ is the Eddington mass accretion rate corresponding to the radiative efficiency of a thin disk (see \autoref{eqn:mdot_edd} and \autoref{eqn:radiative_efficiency_thin}).  Thin disks below and near the Eddington limit are notoriously difficult to simulate, due to difficulties resolving the disk scale height.  However, the additional magnetic pressure of the MAD state helps to inflate even moderately sub-Eddington disks (see \autoref{sec:scale_height}), making this problem computationally tractable \citep[see e.g.,][]{Sadowski2016}.  

Using a mesh-based, finite-difference method in a stationary Kerr space-time, {\sc koral} solves the conservation equations of GRMHD, with the addition of radiative heating, cooling, and plasma coupling.  Modeled radiative processes include synchrotron radiation, opacities from electron scattering, free-free and bound-free emission/absorption from the \cite{Sutherland-Dopita1993}  model, and Compton scattering.  While ideal GRMHD simulations without radiation are rescalable to different masses and accretion rates, the inclusion of radiative processes sets absolute physical scales and necessitates individual simulations for each combination of $M$, $a_*$, and $f_\mathrm{Edd}$.

\begin{figure*}
  \centering
  \includegraphics[width=\textwidth]{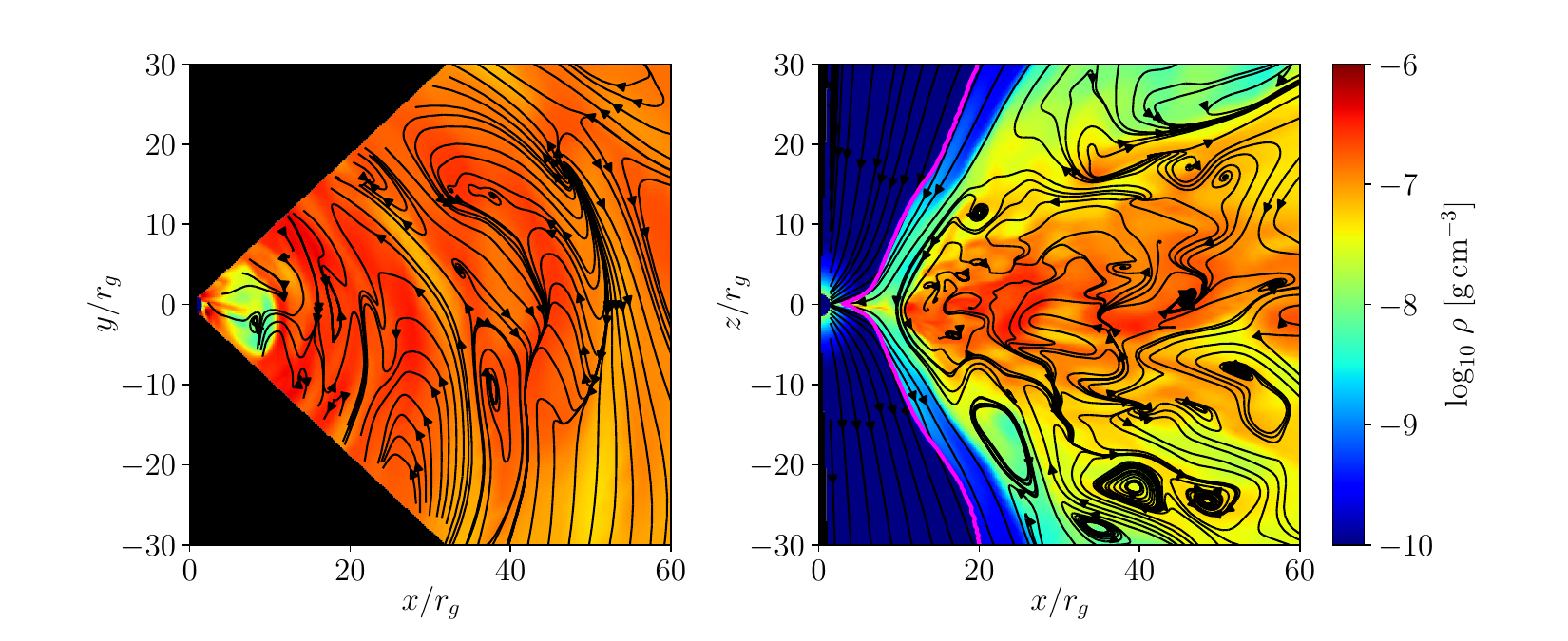}
  \includegraphics[width=\textwidth]{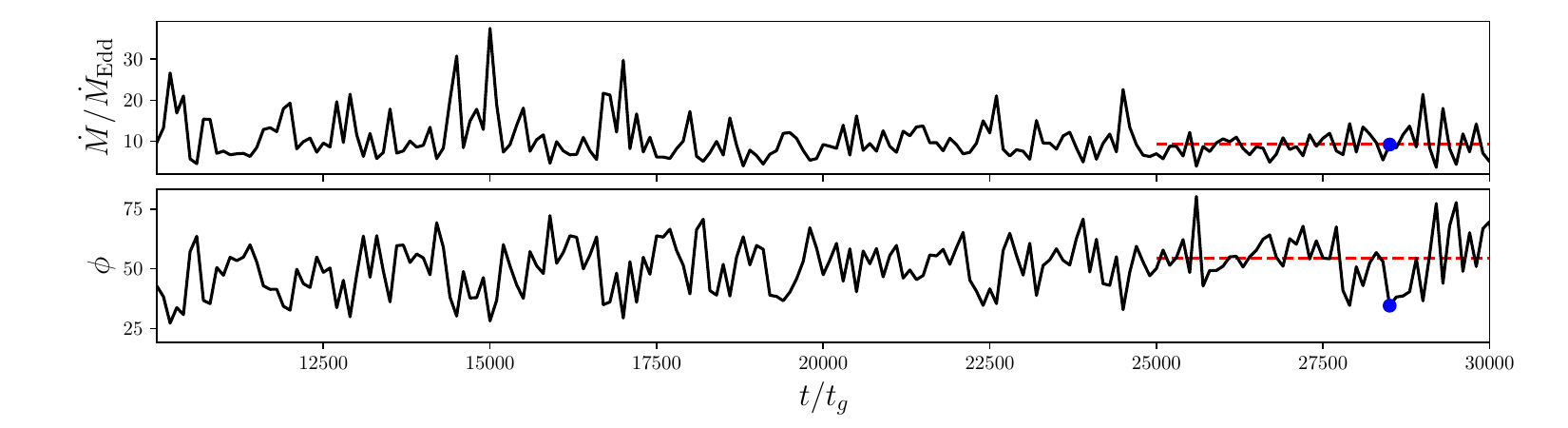}
  \caption{Here we visualize the disk structure and time evolution of the $M=10^4\,M_\odot$, $a_*=0.9$, $f_{\rm{Edd}}=9.3$ model. In the upper two panels, we plot the gas density (color) and magnetic field (black streamlines) within and perpendicular to the disk midplane respectively.  On the right, we plot with a magenta curve the $\sigma \equiv B^2/4\pi \rho =1$ contour, a common definition of the jet boundary.  This snapshot, which corresponds to time $t=28,500\,t_g$, features a flux eruption event, a transient low-density bubble near the horizon.  In the lower panels, we plot the Eddington ratio $f_\mathrm{Edd} = \dot{M}/\dot{M}_\mathrm{Edd}$ and the magnetic flux parameter $\phi$ as a function of time for this model, demonstrating stability for our period of interest, demarcated by the red horizontal lines.  The time corresponding to the snapshot in the upper panesl is marked with a blue circle.}
  \label{fig:visualization}
\end{figure*}

Each simulation is initialized as a torus of gas in hydrostatic equilibrium threaded by a large-scale poloidal magnetic field, either perfectly aligned or anti-aligned with the BH spin axis.  To limit computational expense, but still allow non-axisymetric structures that commonly arise in MAD disks, we simulate a periodic $\pi/2$ wedge in azimuth.  From the torus initial conditions, the magnetorotational instability naturally develops to allow the plasma to lose angular momentum and accrete onto the BH, advecting along with it magnetic field which saturates at the MAD state.  One example is shown in \autoref{fig:visualization}, where in the upper panels we visualize the density and magnetic field lines of the $M=10^4 \ M_\odot$, $a_*=0.9$, $f_\mathrm{Edd}=9.3$ model in the plane and in a perpendicular slice respectively.  The BH has accumulated a significant poloidal magnetic field, and turbulent eddies are evident in the disk.  A flux eruption event characteristic of the MAD state, the low-density bubble near the horizon, is visible during this snapshot.

Throughout this work, we use gravitational units to describe physical parameters. For distance we use the gravitational radius $r_g \equiv GM/c^2$ and for time we use the gravitational time $t_g \equiv GM/c^3$. We set $G = c = 1$, so the above relations would be equivalent to $r_g = t_g = M$. We restore $G$ and $c$ in cases where it helps to keep track of units.  Each of the 38 models was run for a total time of $30000\,t_g$. Summary statistics are given in \autoref{tab:modelinfo} and correspond to averages over the final $5000\,t_g$ of the run when we expect each simulation to be most nearly in steady state.

\section{Results}
\label{sec:results}

\subsection{Magnetization}

The dimensionless magnetization parameter $\phi(t)$ at time $t$ is defined by \citep[][]{Tchekhovskoy+2011}, 
\begin{equation}
    \phi(t) = \frac{\sqrt{4\pi}}{2\sqrt{\dot{M}(t)}} \int_\vartheta \int_\varphi \left|B^r\right|_{r=r_{\rm H}} \;\sqrt{-g}\;\mathrm{d}\vartheta\; \mathrm{d}\varphi,
    \label{eq::madparam}
\end{equation}
where $B^r$ is the radial component of the magnetic field, $g$ is the metric determinant, $\dot{M}(t)$ is the BH accretion rate, and the integral is evaluated at the BH horizon.  MAD systems are characterized by a value of $\phi$ that has saturated at a spin-dependent value of $\sim 30-50$ \citep{Tchekhovskoy+2011,Tchekhovskoy+2012,Narayan+2022}, as is the case for the example plotted in Figure \ref{fig:visualization}.  The value of $\phi$ tends to decrease during a flux eruption event; note that our example snapshot visualized in \autoref{fig:visualization} coincides with a local minimum in $\phi$.  Although both $\dot{M}$ and $\phi$ are time variable, we assign a single value to each simulation by averaging each quantity over the time period $t=25000t_g - 30000t_g$. These are the values listed in \autoref{tab:modelinfo}.

\begin{figure*}
  \centering
  \includegraphics[width=\textwidth]{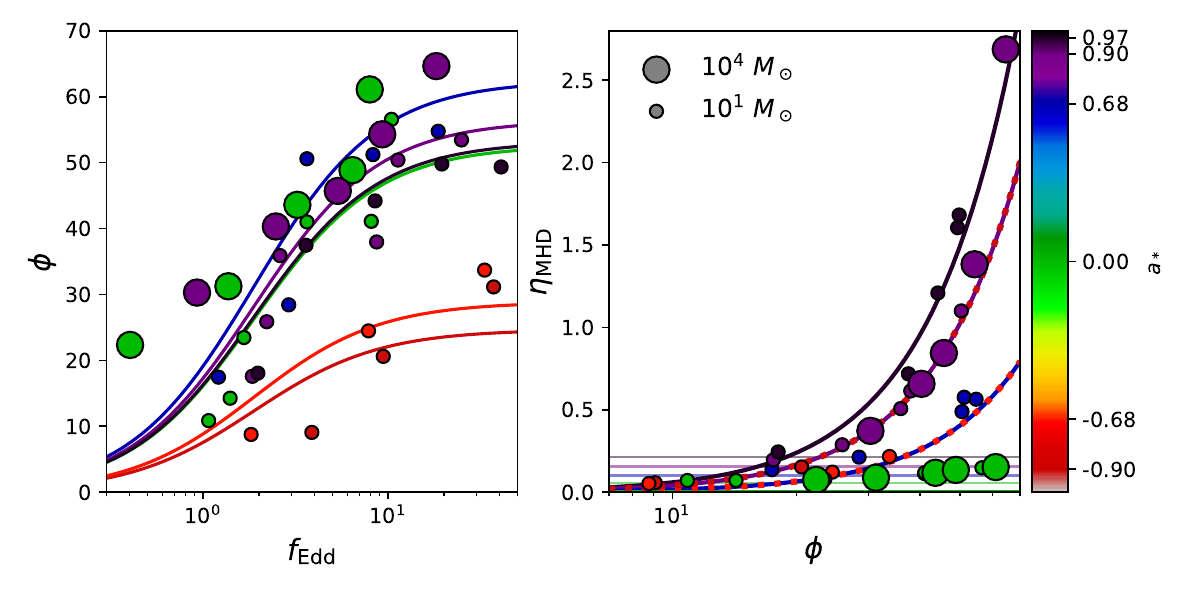}
  \caption{{\it Left:} Magnetic flux parameter $\phi$ as a function of Eddington ratio $f_{\rm Edd}$, where color encodes different values of the BH spin $a_*$.  For each spin sampled by our simulation library, we plot our fitting function (\autoref{eqn:phi}) in the appropriate color. {\it Right:} MHD energy outflow efficiency $\eta_{\rm MHD}$ as a function of magnetic flux parameter for each of our models.  For each spin sampled by our simulation library, we plot the BZ prediction $\eta_{\rm EM}$ (\autoref{eqn:BZ}) as colored lines. The agreement is excellent, implying that a BZ-like electromagnetic jet dominates the outflow energy in most of the simulations, except for $a_*=0$, which features a weaker hydrodynamic outflow. As a point of reference, we plot the radiative efficiencies of thin disks with $a_* \in \{0,0.68,0.9,0.97\}$ as horizontal lines.}
  \label{fig:phimdot}
\end{figure*}

In the left panel of \autoref{fig:phimdot}, we show the values of $\phi$ obtained from our 38 simulations, both as a function of the Eddington ratio $f_\mathrm{Edd}$ and the BH spin $a_*$.  Different spins are encoded in different colors, and different masses are encoded by symbol size.  At large Eddington ratios, the simulations approach spin-dependent values similar to those found in pure GRMHD simulations of MADs \citep{Tchekhovskoy+2012,Narayan+2022}. However, $\phi$ decreases as $f_\mathrm{Edd}$ decreases.  Interestingly, simulations with $f_\mathrm{Edd}=1$ remain substantially magnetized, with $\phi$ values typically about a third of the limiting value for $f_{\rm Edd}\gg1$.  As we explore in \autoref{sec:scale_height}, this trend can be explained by increased pressure scale height as Eddington ratio increases, allowing the disk to confine stronger magnetic fields.

We model the behavior shown in the simulation data by fitting the following function:
\begin{equation}
    \phi(a_*,f_\mathrm{Edd}) = \phi_\mathrm{MAD}(a_*)\frac{(f_\mathrm{Edd}/f_c)^\alpha}{1+(f_\mathrm{Edd}/f_c)^\alpha}, \label{eqn:phi}
\end{equation}
where $f_c$ is a critical Eddington ratio determining the mid-point of the transition, and $\alpha$ is a free parameter determining the rapidity of the evolution around $f_c$. The function $\phi_\mathrm{MAD}(a_*)$ is the saturated value of $\phi$ found in non-radiative MAD simulations. We use the approximation given in \citet{Narayan+2022}, 
\begin{equation}
    \phi_\mathrm{MAD}(a_*) = 52.6 + 34a_*  - 14.9a^2_* - 20.2a^3_*.
    \label{eqn:phi_mad}
\end{equation}
By construction, in \autoref{eqn:phi}, $\phi \to 0$ as $f_\mathrm{Edd} \to 0$ and $\phi \to \phi_\mathrm{MAD}(a_*)$ as $f_\mathrm{Edd} \to \infty$.  Via least-squares fitting, we arrive at $\alpha=1.29$ and $f_c=1.88$.  The spin-dependent $\phi(a_*,f_\mathrm{Edd})$ curves are plotted in the background of \autoref{fig:phimdot}, and describe the main trends fairly well.  We intentionally transition $\phi \to 0$ as $f_\mathrm{Edd} \to 0$ to connect to the thin disk solution, but we caution that the shape and rapidity of this transition may be sensitive to our poor sampling of the $f_\mathrm{Edd} \lesssim 1$ regime.  We note that the GRRMHD simulations of both \citet{Liska+2022} and \citet{Curd&Narayan2023} produced $\phi \sim 30$ for $f_\mathrm{Edd} \sim 0.3$, which our fitting function would underestimate.

\subsection{Jet Efficiency}

The electromagnetic jet efficiency $\eta_\mathrm{EM} = P_\mathrm{jet} / \dot{M}c^2$ can be calculated analytically given $a_*$ and $\phi$.  For small to moderate values of spin, $\eta_{\rm EM} \propto a_*^2\phi^2$ \citep{Blandford&Znajek1977}, but for spin values up to and including $a_*=1$, the following expression including higher order correction factors is more accurate \citep{Tchekhovskoy+2010,Pan&Yu2015}:
\begin{equation}
    \eta_\mathrm{EM} = \frac{\kappa}{4\pi}\phi^2\Omega^2_{\rm H}\left[1 + 1.38\Omega^2_{\rm H} - 9.2\Omega^4_{\rm H}\right],
\label{eqn:BZ}
\end{equation}

\noindent where 

\begin{equation}
    \Omega_{\rm H}\equiv \frac{|a_*|}{2r_{\rm H}}=\frac{|a_*|}{2(1+\sqrt{1-a_*^2})}
    \label{eqn:Omega_H}
\end{equation}
is the angular velocity of the horizon and $\kappa$ is a constant dependent on the initial field geometry, for which we adopt $\kappa = 0.05$. 

In the right panel of \autoref{fig:phimdot}, we plot the MHD energy outflow efficiency $\eta_\mathrm{MHD}$ as a function of magnetization, with spin once again encoded in color and mass encoded in symbol size.  Note that unlike $\eta_\mathrm{EM}$ predicted by \autoref{eqn:BZ} this quantity also includes the hydrodynamic energy flux.  The colored curves correspond to the fitting function \autoref{eqn:BZ} for each spin sampled by our simulation suite. The data points are from the simulations, where we have computed the mass and energy fluxes at a radius of $5 \ r_g$ since numerical floors cause inaccuracies closer to the horizon \citep[consistent with previous work][]{Lowell+2023}.  Radiative flux is neglected (which is again affected by floors, particularly in the jet region), but this introduces only a small error since the radiation contribution near the BH tends to be small.  

Despite the wide range of mass, spin and accretion rate considered in the right panel of \autoref{fig:phimdot}, we find that the fitting function \autoref{eqn:BZ} performs remarkably well, implying that the BZ mechanism dominates the jet physics in MAD super-Eddington accretion flows. Note that at $a_*=0$, the BZ prediction is identically 0 because the BH has no spin energy. However, the simulations still give $\eta_\mathrm{MHD}>0$. In these models, the outflowing energy is from the accretion disk, presumably in a hydrodynamic wind. As a point of reference, we plot the radiative efficiencies of thin disks with $a_* \in \{0,0.68,0.9,0.97 \}$ as colored horizontal lines. The MHD outflow from the $a_*=0$ simulation is similar in energetic output to an equivalent thin disk's radiative output.  Meanwhile, the radiative efficiency of a thin disk around a maximally spinning black hole can be easily be exceeded with enough spin and magnetic flux.

\subsection{Spin Evolution}
\label{sec:spin}

\begin{figure*}
  \centering
  \includegraphics[width=\textwidth]{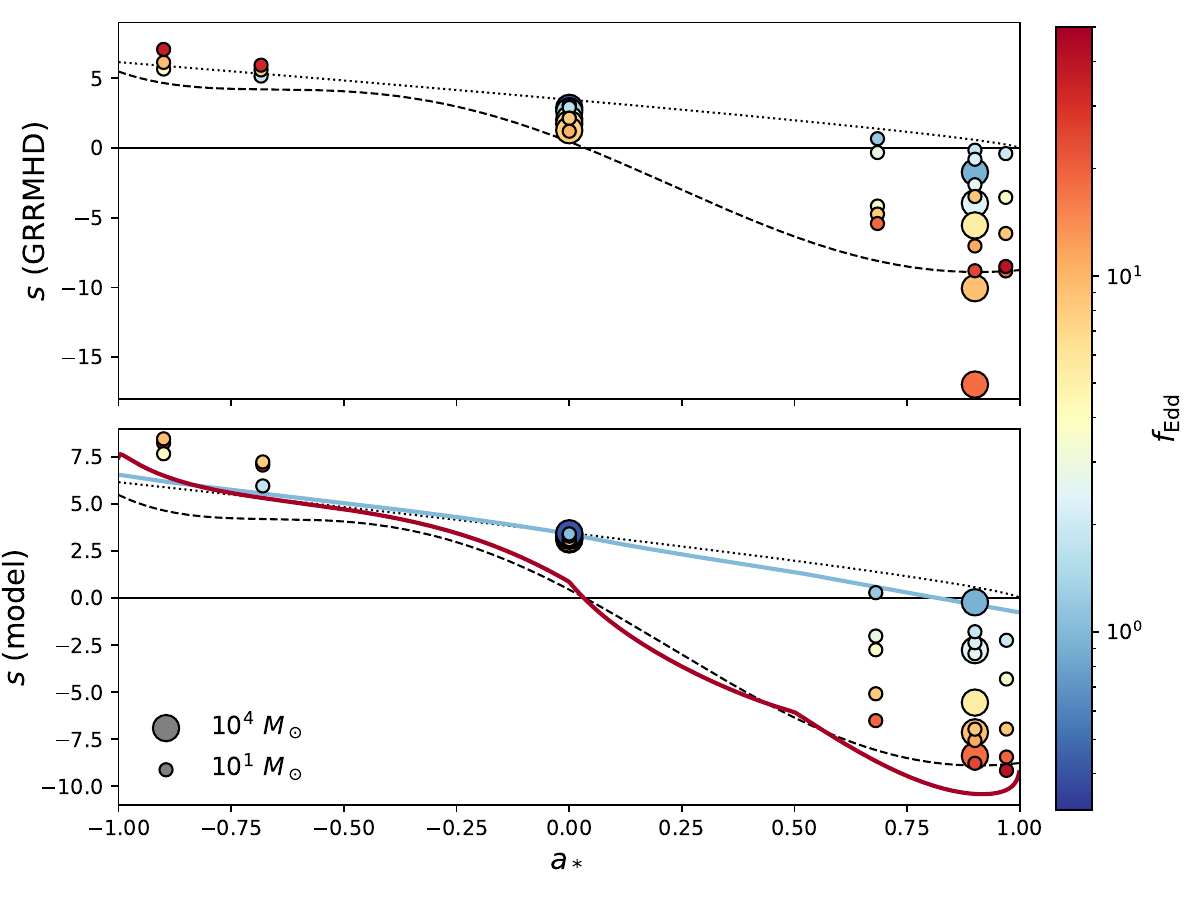}
  \caption{Spinup parameter $s$ as a function of BH spin $a_*$, with Eddington ratio $f_{\rm Edd}$ encoded in the color.  Values computed directly from our GRMHD simulations are plotted in the upper panel, and the predictions of our fitting functions (\autoref{eqn:spinup}) are shown in the lower panel.  At the lowest accretion rates, models approximately match the prediction for a razor-thin disk (\autoref{eqn:s_thin}), shown as the dotted line.  At the highest accretion rates, prograde and zero-spin models approach the curve found for pure GRMHD models (\autoref{eqn:s_mad}), plotted as a dashed line.  We plot our model predictions for $s$ for $f_\mathrm{Edd}=1$ and $f_\mathrm{Edd} \to \infty$ with light blue and dark red curves respectively.}
  \label{fig:spindown}
\end{figure*}

Since the BZ mechanism extracts spin energy from the BH, this can result in astrophysically significant spin evolution of an accreting BH, which we study here.  We describe the evolution in terms of a dimensionless spin-up parameter \citep{Gammie+2004,Shapiro2005}, 
\begin{equation}
    s = \frac{da_*}{dt}\frac{M}{\dot{M}} = l - 2 a_* e,
    \label{eqn:s_def}
\end{equation}
where $l$ is the inward specific angular momentum flux and $e$ is the inward specific energy flux, each of which we measure at a radius of $5 \ r_g$.  Spinup as a function of $a_*$ computed from our GRRMHD simulations is shown in the upper panel of \autoref{fig:spindown}, where the color encodes different Eddington ratios and the symbol size encodes different masses.  The thin disk solution, which always pushes the BH towards maximal prograde spin ($a_*\to 1$), is shown as a dotted line \citep{Novikov&Thorne1973,Moderski&Sikora1996}.  A fitting function which we presented in previous work for MAD GRMHD ($f_{\rm Edd} \ll 1$) models \citep{Narayan+2022}  is shown as a dashed line and is given by
\begin{equation}
    \begin{aligned}
    s_\mathrm{MAD}(a_*) = & 0.45 - 12.53 a_* -7.80 a_*^2 + 9.44 a_*^3 \\
    &+5.71 a_*^4 - 4.03 a_*^5.
    \label{eqn:s_mad}
    \end{aligned}
\end{equation}

The simulated GRRMHD models generally transition from the thin disk solution to the MAD GRMHD solution as the Eddington ratio increases (blue to red colors in \autoref{fig:spindown}). This is not unexpected, since highly super-Eddington disks are geometrically very thick and are highly advection-dominated \citep{Abramowicz+1988} and therefore closely resemble the low-$f_{\rm Edd}$ hot accretion flows studied in \citet{Narayan+2022}. Retrograde models do not follow this trend, however, in fact spinning up more rapidly than the thin disk solution.  These models overshoot the thin disk curve because both the BZ mechanism and accretion of oppositely rotating material torque the BH towards $a_*=0$.\footnote{As Eddington ratio increases, the disk dynamics evolve from the thin disk solution and the hydrodynamic torques become weaker (see \autoref{sec:fluxes}). At the same time, the magnetization increases, so the electromagnetic torque becomes {\it stronger}.  Whether or not a retrograde disk spins up faster or slower than a thin disk depends on the balance between these effects.}

\citet{Lowell+2023} built a semi-analytic model to understand spin evolution in non-radiative MAD systems based on the spin evolution equations appropriate for a disk-plus-jet system introduced in \citet{Moderski&Sikora1996}.  In this model, the spinup parameter is explicitly split up into hydrodynamic spinup by the accretion disk gas and spindown via a jet powered by the BZ mechanism.  The spinup parameter is then expressed as
\begin{equation}
    s = s_\mathrm{HD} + s_\mathrm{EM} \label{eqn:spinup_edd},
\end{equation}
where
\begin{align}
    s_\mathrm{HD} = l_\mathrm{HD} - 2 a_* e_\mathrm{HD},
\end{align}

\noindent and

\begin{align}
    s_\mathrm{EM} =\mathrm{sign}(a_*)\,\eta_\mathrm{EM} \left( \frac{1}{k \Omega_H} - 2 a_* \right).
\end{align}

We detail the calculation and modeling of $s_\mathrm{HD}$ from $l_\mathrm{HD}$ (the hydrodynamic specific angular momentum flux) and $e_\mathrm{HD}$ (the hydrodynamic specific energy flux) in \autoref{sec:fluxes}.  As explained there, we develop a fitting function for $s_\mathrm{HD}$ given by \autoref{eqn:sHD} that smoothly interpolates between the thin disk solution as $f_\mathrm{Edd} \to 0$ and non-radiative GRMHD results as $f_\mathrm{Edd} \to \infty$.  Meanwhile, the electromagnetic component $s_\mathrm{EM}$ depends on $\eta_\mathrm{EM}$ and the parameter $k$, which is the ratio of the angular frequency of field lines relative to that of the BH.  We estimate $\eta_\mathrm{EM}$ as a function of $a_*$ and $f_\mathrm{Edd}$ by combining \autoref{eqn:BZ} and \autoref{eqn:phi}.  For $k$, we adopt the following fit from the non-radiative GRMHD simulations of \citet{Lowell+2023}:
\begin{equation}
    k(a_*) = \begin{cases}
        0.23, & a_* < 0 \\
        \mathrm{min}(0.1+0.5a_*,0.35), & a_* > 0
    \end{cases}
    \label{eqn:k}
\end{equation}

\noindent this gives $k$ slightly less than the \citet{Blandford&Znajek1977} monopole value of 0.5, which broadly agrees with other simulations in the literature \citep{McKinney+2012,Penna+2013b,Chael+2023}.

As one final modification to allow our model to support hot accretion flows, we make the following adjustment:

\begin{equation}
    s= \begin{cases}
        s_\mathrm{HD} + s_\mathrm{EM} & f_\mathrm{Edd} > f_c \\
        s_\mathrm{MAD} & f_\mathrm{Edd} \leq f_c
    \end{cases}
    \label{eqn:spinup}
\end{equation}

\noindent where $f_c$ is a critical Eddington ratio below which the accretion flow should transition to the radiatively inefficient hot accretion mode \citep[][]{Narayan-Yi1994,Narayan-Yi1995,Abramowicz+1995}.  Following previous efforts to model the evolution of black hole populations, we adopt $f_c=3\times 10^{-2}$ \citep{Merloni&Heinz2008,Volonteri+2013}.  The exact Eddington ratio at which this transition occurs is poorly constrained and unlikely to be a sharp transition \citep{Cho&Narayan2022}.  Different values of $f_c$ may be adopted without qualitatively changing our formulae.

Our final result for the spinup parameter $s$ (\autoref{eqn:spinup}) can thus be obtained from just two parameters ($a_*$ and $f_\mathrm{Edd}$) by inserting our fitting functions for $\phi(a_*,f_\mathrm{Edd})$ (\autoref{eqn:phi}), $s_\mathrm{HD}(a_*,f_\mathrm{Edd})$ (\autoref{eqn:sHD}), and $\eta_\mathrm{EM}(a_*,\phi)$ (\autoref{eqn:BZ}).  As constructed, \autoref{eqn:spinup} can be applied to all physical values of $a_* \in [-1,1]$ and $f_\mathrm{Edd}\in (0,\infty)$.  

The model predictions from  \autoref{eqn:spinup} are shown in the bottom panel of \autoref{fig:spindown}.  The model captures the behavior seen in the simulations (upper panel) exceptionally well, especially for spinning BHs.  For $a_*=0$, it underestimates the evolution of $s$ with $f_\mathrm{Edd}$.  We speculate that this may be due to the exclusion of angular momentum loss due to hydrodynamic wind, evident in \autoref{fig:phimdot}.  In light blue, we plot the model's prediction for $s$ when $f_\mathrm{Edd}=1$.  It is quite similar to the thin disk solution, but has a root, which corresponds to an equilibrium value of $a_*$ for fixed $f_\mathrm{Edd}$, at $a_{*,\rm eq}\approx 0.8$ instead of 1.  In red, we plot the limit as $f_\mathrm{Edd} \to \infty$.  It follows the non-radiative GRMHD fitting function well, with minor deviations in the retrograde regime.  This curve exhibits two kinks originating from the piece-wise nature of \autoref{eqn:k}.  As $f_\mathrm{Edd} \to f_c$, $s$ is well-approximated by the thin disk solution (dotted black line) by construction.  In any case, the key result from the red line is that, as $f_{\rm Edd} \to \infty$, the equilibrium spin (where $s=0$) approaches $a_{*,\rm eq}\approx0$.

\begin{figure}
  \centering
  \includegraphics[width=0.5\textwidth]{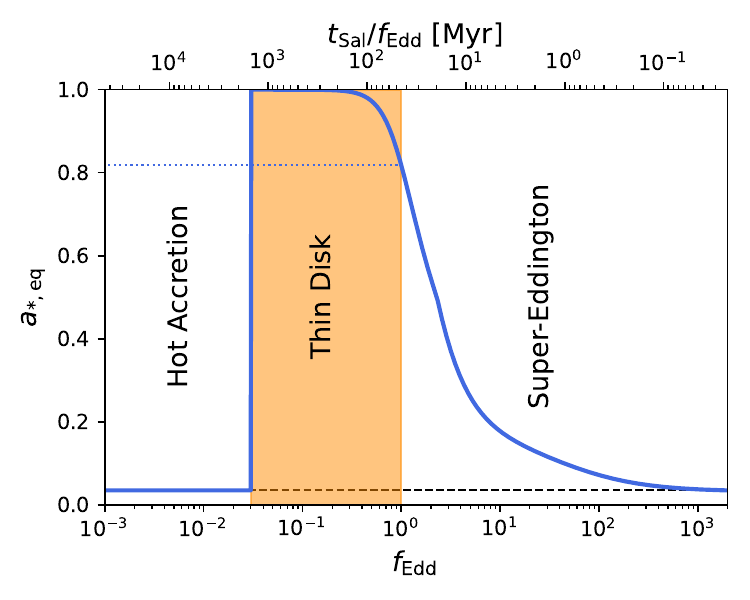}
  \caption{Equilibrium spin $a_{*,\rm eq}$ as a function of Eddington ratio $f_{\rm Edd}$ using our model (\autoref{eqn:spinup}).  Systems with $f_\mathrm{Edd}=1$ reach equilibrium at $a_{*,\mathrm{eq}} \approx 0.8$, while those with a factor of a few smaller $f_{\rm Edd}$ equilibrate near $a_{*,\mathrm{eq}} \approx 1$.  Systems with both $f_{\rm Edd}\ll1$ and $f_{\rm Edd}\gg1$ reach equilibrium near $a_{*,\rm eq}\approx0$. In the upper $x$-axis, we plot $t_\mathrm{Sal}/f_\mathrm{Edd}$, the timescale over which both mass and spin evolve and thus the minimum timescale required to reach spin equilibrium.}
  \label{fig:equilibrium}
\end{figure}

In \autoref{fig:equilibrium}, we plot the equilibrium spin $a_{*,\rm eq}$ as a function of Eddington ratio, found by taking \autoref{eqn:spinup} and solving the condition $s=0$ at fixed $f_\mathrm{Edd}$.  We demarcate three different physical regimes: (i) hot accretion for $f_\mathrm{Edd} < f_c$, (ii) what is classically modeled as a thin disk for $f_c < f_\mathrm{Edd} < 1$, and (iii) super-Eddington accretion for $f_\mathrm{Edd} >1$.  In reality, $s$ and $a_{*,\mathrm{eq}}$ should evolve more gradually around $f_{\rm Edd} \approx f_c$, but we lack a detailed understanding of this transition and are unable to model it more realistically in this work.

Our model permits the existence of BHs with a stable $a_{*,\rm eq} \approx 1$ for Eddington ratios in the range $f_{\rm Edd} \sim 0.03 - 0.3$, but $a_{*,\mathrm{eq}}$ begins to decline above $f_\mathrm{Edd} \approx 0.3$ and approaches 0 as the accretion rate becomes highly super-Eddington.  The limiting equilibrium spin for extremely large values of $f_\mathrm{Edd}$ is $a_*=0.035$, as in the hot accretion regime \citep{Narayan+2022,Lowell+2023}, but note that this exact value is not very accurate and depends on the details of how spin-down is modeled.  On the upper $x$-axis, we plot the evolutionary timescale of both mass and spin for a given $f_\mathrm{Edd}$, given by $t_\mathrm{Sal}/f_\mathrm{Edd}$ where
\begin{equation}
    t_\mathrm{Sal} = \frac{\epsilon \sigma_T c}{4 \pi G m_p} = \epsilon \times 450 \ \mathrm{Myr}
\end{equation}
is called the Salpeter timescale, where $\sigma_T$ is the Thomson cross-section and $m_p$ is the proton mass.  For the convenience of defining a spin-independent $t_\mathrm{Sal}$, we adopt a fiducial value of $\epsilon=0.1$ for its definition, such that $t_\mathrm{Sal} = 45 \ \mathrm{Myr}$.  Since mass and spin evolve on the same time-scale, a BH must accrete a significant fraction of its own mass to reach equilibrium spin\footnote{However, note that $s$ measures the ratio of the spin evolution rate to the mass evolution rate. Hence for values of $|s|$ approaching 10, spin evolves 10 times faster than mass.}.  In the hot accretion regime, this would occur on timescales easily exceeding the age of the universe, and thus such BHs will not naturally reach the equilibrium spin value through the BZ process \citep[although noticeable evolution is still possible;][]{Narayan+2022}.  However, BHs which accrete continuously near or above the Eddington limit can reach their equilibrium spins in less than (sometimes very much less than) a Hubble time.  Interestingly, such continuous and rapid assembly is invoked to explain the existence of massive quasars at $z \gtrsim 6$ \citep[e.g.,][]{Fan+2003,Banados+2018,Wang+2021,Bogdan+2023}, which have accumulated masses up to $10^{10} \ M_\odot$ when the Universe was approximately 1 Gyr old.  

\section{Discussion and Conclusions}
\label{sec:conclusion}

In this letter we presented a suite of GRRMHD simulations of radiative MAD accretion disks around BHs. The simulations cover a range of BH spins $a_*$ from $+0.97$ to $-0.9$, and Eddington ratios $f_{\rm Edd}$ from 0.4 to 40. We find two key qualitative results.

First, radiative disks in the MAD state around spinning BHs produce powerful jets as efficiently as the better-studied non-radiative disks (which are found in systems with $f_{\rm Edd} \ll 1$), and the power in the jet comes similarly from the BZ mechanism (see the right panel of \autoref{fig:phimdot}).

Second, the saturated magnetic flux $\phi$ depends not only on the BH spin (as already known for non-radiative MAD models) but also on the Eddington ratio (see the left panel of \autoref{fig:phimdot}). As a result, radiative disks with $f_{\rm Edd} \lesssim 0.3$ behave roughly like the standard thin accretion disk model, but systems with $f_{\rm Edd} \gg 1$ are very different and closely resemble non-radiative models (see \autoref{fig:equilibrium}). In particular, when $f_{\rm Edd}\gg1$, the accreting BHs spin-down rapidly toward an equilibrium $a_*\approx 0$.

At a quantitative level, using the above suite of MAD GRRMHD simulations we have devised fitting functions which can be used to estimate magnetization $\phi$ (\autoref{eqn:phi}), jet feedback efficiency $\eta$ (\autoref{eqn:BZ}), and spin evolution $s$ (\autoref{eqn:spinup}), as a function of spin and Eddington ratio.  Spindown via the BZ mechanism grows more efficient as Eddington ratio increases, but is already noticeable at $f_\mathrm{Edd} \approx 1$, where the equilibrium spin is $a_*=0.8$.  This has important implications for feedback and spin-evolution of BHs in the near-Eddington to super-Eddington regime, such as flux-limited samples of AGN, rapidly assembling seeds in the early universe, and collapsar BHs.

\begin{figure*}
  \centering
  \includegraphics[width=\textwidth]{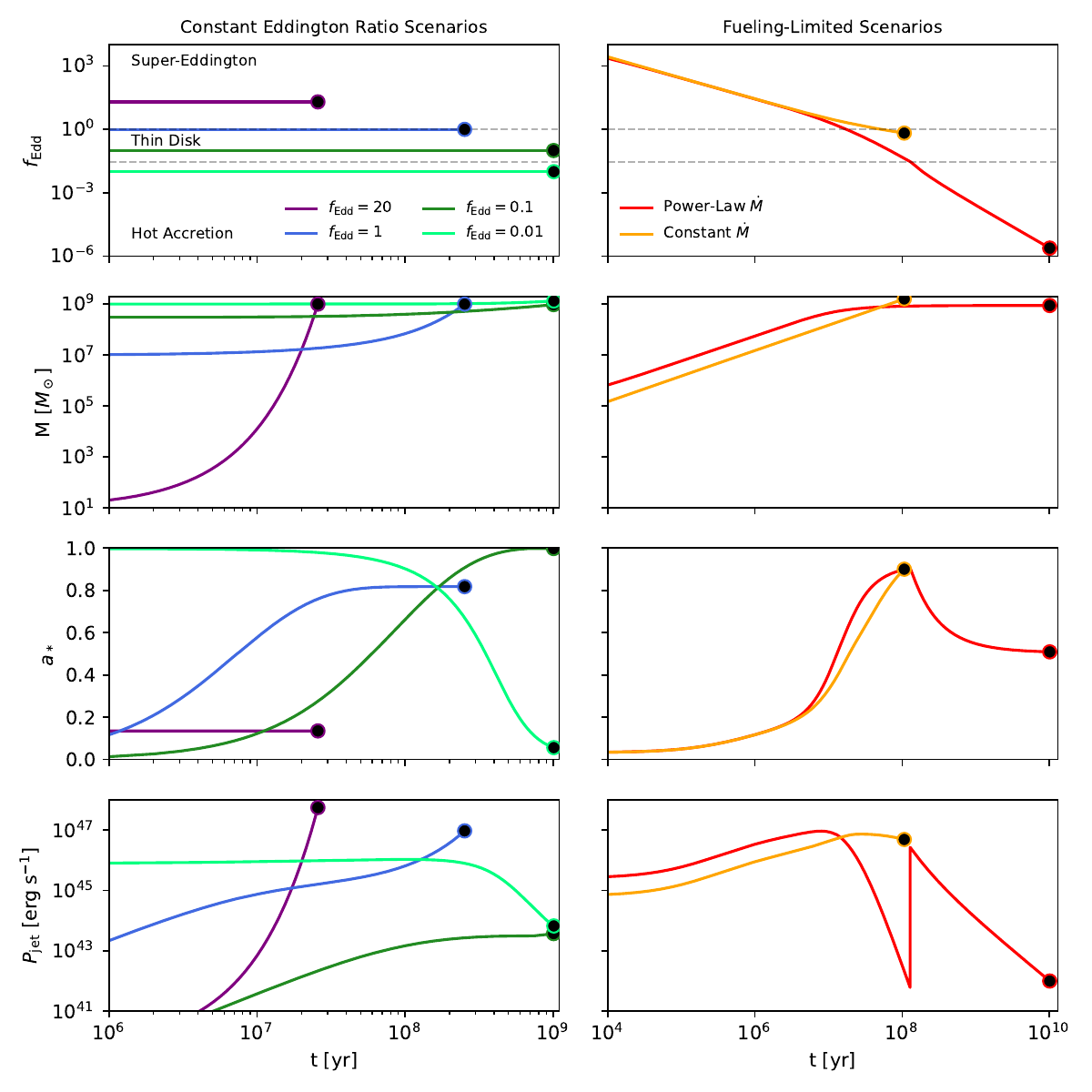}
  \caption{Example evolutionary pathways of BH mass and spin computed using the fitting functions derived in this work. In each column, the panels show from top to bottom the Eddington ratio $f_\mathrm{Edd}$, the BH mass $M$, BH spin $a_*$, and the jet power $P_\mathrm{jet}$ as a function of time.  Each pathway is tuned to produce $M \approx 10^9 \ M_\odot$ at the final time.  {\it Left: } Constant Eddington ratio scenarios, each of which reaches a distinct equilibrium spin from which their recent cosmically averaged $f_\mathrm{Edd}$ could be inferred.  {\it Right: } Fueling-limited scenarios where we prescribe $\dot{M}$ as a function of time.  The Constant $\dot{M}$ scenario is tuned to match the \citet{Wang+2021} quasar found when the Universe was only 670 Myr old ($z=7.642$).  The Power-Law $\dot{M}$ scenario has a prescribed time-variable accretion rate, $\dot{M} \propto (1+(t/10^7 \ \mathrm{yr})^{2})^{-1}$ \citep[motivated by][]{Hopkins+2006b,Hopkins+2006}, and might represent the history of a currently low-Eddington rate SMBH at the center of a galaxy cluster such as Messier 87.}
  \label{fig:evolution_gallery}
\end{figure*}

In \autoref{fig:evolution_gallery}, we plot evolutionary tracks for a selection of cosmologically motivated scenarios, each of which results in a BH with $M \approx 10^9 \ M_\odot$.  In each case, we have integrated \autoref{eqn:spinup} using a standard Runge-Kutta-Fehlberg 4(5) integrator with adaptive step-sizing.  For these examples, we make an important assumption that the accretion disk and BH angular momentum axes are always perfectly aligned, which need not generally be the case.  Variations in disk tilt over cosmic time are an uncertainty that can lead to substantial differences in spin evolution, leading to lower spins if the angular momenta of material is more randomized \citep{King+2008,Berti&Volonteri2008}.  In the left column of \autoref{fig:evolution_gallery}, we plot evolutionary scenarios with different fixed $f_\mathrm{Edd}$ values shown as different colors.  For $f_\mathrm{Edd}=20, ~1, ~0.1, ~0.01$, we initialize our BHs with $M=10, ~10^7, ~3\times10^8, ~10^9 \ M_\odot$ and $a_*=0, ~0, ~0, ~0.998$, respectively.  In all cases, 1 Gyr is enough for each of the BHs to approach their equilibrium spin (see \autoref{fig:equilibrium}).  These scenarios result in very different spin evolution and feedback as a function of time.  

Both the $f_\mathrm{Edd}=20$ and the $f_\mathrm{Edd}=1$ scenarios result in the accretion of $10^9 \ M_\odot$ of material, but the $f_\mathrm{Edd}=20$ scenario releases a total of $7.8 \times 10^{53} \ \mathrm{erg}$ worth of feedback compared to $5.3 \times 10^{54} \ \mathrm{erg}$ in the $f_\mathrm{Edd}=1$ scenario, a factor of 7 difference.  The reason is that the $f_{\rm Edd}=20$ model reaches a lower equilibrium spin, which results in less efficient jet feedback.  A consequence of this interesting result is that a BH could potentially grow {\it more} efficiently in a super-Eddington state before having its mass supply cut off by excessive jet feedback.  We have assumed a sharp transition between thin and thick accretion flows at an Eddington ratio of $f_c = 3\times10^{-2}$.  Evolving in the thin disk regime, the $f_\mathrm{Edd}=0.1$ model spins {\it up} to maximal spin and cannot power a very efficient jet, since lower Eddington ratio sources maintain weaker magnetization.  On the other hand, the $f_\mathrm{Edd}=0.01$ model evolves in the hot accretion flow regime and spins {\it down} to near zero spin.  

In the right column of \autoref{fig:evolution_gallery}, we plot two different fueling-limited scenarios.  In the ``Constant $\dot{M}$'' model, we envision that a galaxy provides constant $\dot{M}$ that the BH can consume, regardless of the $f_\mathrm{Edd}$ implied.  In this model, we suggestively tune our parameters to match the formation of the \citet{Wang+2021} quasar, which is observed with $f_\mathrm{Edd}=0.67$ and $M=1.6\times 10^9 \ M_\odot$ at $z=7.642$, when the Universe was only 670 Myr old.  After being initialized at $10^4 \ M_\odot$ and $a_*=0$, the BH accumulates mass in the super-Eddington regime as spindown from the BZ mechanism keeps its spin low.  Its spin increases only as $f_\mathrm{Edd} \to 1$, and it reaches an equilibrium spin of $0.9$.  Qualitatively consistent with our predictions for a powerful jet, \citet{Wang+2021} report a relativistic outflow while also suggesting greater incidence of such powerful outflows at high redshift. 

In the second ``Power-Law $\dot{M}$'' model, a $10^5 \ M_\odot$ $a_*=0$ seed initially accretes at $f_\mathrm{Edd}=15,000$, then the accretion rate declines as $\dot{M} \propto (1+(t/10^7 \ \mathrm{yr})^{2})^{-1}$, motivated by \citet{Hopkins+2006b,Hopkins+2006}.  Over the age of the Universe, this BH traverses all three accretion regimes, starting with $a_* \approx 0$ while it is super-Eddington, rising to $a_* \approx 0.9$ in the thin disk regime, then finally declining to $a_* \approx 0.5$ in the hot accretion regime.  It runs out of fuel before it can achieve the equilibrium spin $\approx 0$ for its final $f_{\rm Edd}$.  Ending with $f_\mathrm{Edd} \sim 10^{-6}$ and $M \sim 10^9 \ M_\odot$, this evolutionary track could represent the history of the most massive BHs resolvable on the sky, such as Event Horizon Telescope target Messier 87.  

\autoref{fig:evolution_gallery} illustrates how a BH's assembly history is imprinted on its final spin value, motivating observational spin constraints of supermassive BHs.  For $0.01 \lesssim f_\mathrm{Edd} \lesssim 0.3$, X-ray reflection spectroscopy has been most successful in accumulating large spin samples.  The measured spin values tend to be highly skewed towards $a_* \approx 1$ \citep[see][for a recent review]{Reynolds+2021}, in agreement with the equilibrium spin of a thin accretion disk, as well as the equilibrium spin value suggested by the present work for that range of $f_\mathrm{Edd}$. To complement these thin disk spin constraints, the next-generation Event Horizon Telescope aims to measure spins of dozens of supermassive BHs in the hot accretion ($f_\mathrm{Edd} \ll 1$) regime \citep{Pesce+2022,Ricarte+2023}.  Taking the ``Power-Law $\dot{M}$'' model in \autoref{fig:evolution_gallery} as an example, we would predict typical spin values roughly half-way between 1 and 0 \citep[but recall that these calculations have neglected angular momentum flips and BH-BH mergers, e.g.,][]{Berti&Volonteri2008}. It would be interesting to see what future observations show. Unfortunately, there is no known direct probe of spin in the super-Eddington regime, where we predict equilibrium spins close to 0.  Current probes of spin rely on the existence of a sharp transition in the dynamics of the accreting disk at the innermost stable circular orbit. Such a feature is expected to be present in geometrically thin disks (and is the basis of the X-ray reflection method), but it is washed out in geometrically thick disks such as are found for $f_{\rm Edd}\gg1$ (e.g., this work).

It is worth mentioning that in the present radiative MAD models, as well as others in the literature, roughly $\sim 60$\% of the jet power can be transformed into radiation at large radius \citep{Curd&Narayan2023}.  This can occur because inverse Compton scattering can transform much of the kinetic energy of the jet fluid into highly beamed radiation.  However, we refrain from providing radiative efficiencies from our simulations, because we find that numerical floors in the jet region can artificially inflate the total energy in the jet at large radii. Fortunately, this artificially injected energy simply outflows from the simulation box and does not affect the region of interest.

The analytic formulae devised in this work can be applied to galactic or cosmological scale simulations, conveniently bridging the sub-Eddington and super-Eddington regimes.  When placing these models in an astrophysical context, the most important caveat is the assumption that these systems are magnetically saturated in the MAD state.  Event horizon scale polarimetric imaging the largest black holes on the sky do currently favor MAD models over their SANE counterparts \citep{EHT8,EHT_SgrA_V,Wielgus+2022}, and ab-initio simulations of gas and magnetic field transport onto Sgr A* can indeed naturally produce MAD states \citep{Ressler+2020,Ressler+2023}, but this evidence pertains only to low-Eddington ratio BHs.  Super-Eddington MAD disks can explain jetted tidal disruption events \citep{Tchekhovskoy+2014,Curd+2019}, but these objects are only $\sim$1\% of known TDEs and may not be representative of the typical super-Eddington disk.  Future observational and theoretical developments to test the robustness of the MAD state would help validate the modeling performed here.  Furthermore, our simulations are limited to $M=10 \ M_\odot$ and $M=10^4 \ M_\odot$, and \autoref{fig:phimdot} hints at a possible trend with mass.  We do not expect our results to be very sensitive to BH mass on physical grounds, but this should be verified in future work in the context of varying the metallicity as well.

\section{Acknowledgments}
This work was supported in part by NSF grants AST1816420 and OISE-1743747, and by the Black Hole Initiative at Harvard University, made possible through the support of grants from the Gordon and Betty Moore Foundation and the John Templeton Foundation. The opinions expressed in this publication are those of the author(s) and do not necessarily reflect the views of the Moore or Templeton Foundations.

\software{{\sc koral} \citep{Sadowski+2013,Sadowski+2014}, Matplotlib \citep{matplotlib}, SciPy \citep{scipy}, NumPy \citep{numpy}}

\section{Data Availability}

Most plotted values can be downloaded from data files that accompany this publication.  In addition, we provide a Python script including the equations presented in this work, as well as the integrator that was used to produce \autoref{fig:equilibrium} and \autoref{fig:evolution_gallery}.

\bibliography{ms}

\appendix

\section{Additional GRRMHD Details}
\label{sec:grrmhd}

Using the finite-difference method in a fixed, Kerr spacetime, {\sc koral} solves the conservation equations: 
\begin{align}
  (\rho u^\mu)_{;\mu} &= 0, \label{eq:consrho} \\
  (T^\mu_{\ \nu})_{;\mu} &= G_\nu, \label{eq:consT} \\
  (R^\mu_{\ \nu})_{;\mu} &= -G_\nu, \label{eq:consR} \\
  (nu^\mu_R)_{;\mu} &= \dot{n}, \label{eq:ndot} 
\end{align}
where $\rho$ is the gas density in the comoving fluid frame, $u^\mu$ are the components of the gas four-velocity as measured in the ``lab frame'', $T^\mu_{\ \nu}$ is the MHD stress-energy tensor in the ``lab frame'':
\begin{equation}
  T^\mu_{\ \nu} = (\rho + u_g+ p_g + b^2)u^\mu u_\nu + (p_g + \dfrac{1}{2}b^2)\delta^\mu_{\ \nu} - b^\mu b_\nu,
\end{equation}
$R^\mu_{\ \nu}$ is the stress-energy tensor of radiation, $G_\nu$ is the radiative four-force which describes the interaction between gas and radiation \citep{Sadowski+2014}, and $n$ is the photon number density. Here $u_g$ and $p_g=(\gamma_g - 1)u_g$ are the internal energy and pressure of the gas in the comoving frame, and $b^\mu$ is the magnetic field four-vector which is evolved following the ideal MHD induction equation \citep{Gammie+2003}. For fitting purposes, it is useful to write the MHD stress-energy tensor in terms of hydrodynamic (HD) and electromagnetic (EM) components
\begin{equation} \label{eq:TmunuHD}
  T^\mu_{\ \nu, {\rm{HD}}} = (\rho + u_g+ p_g)u^\mu u_\nu + p_g\delta^\mu_{\ \nu}
\end{equation}
and
\begin{equation}
  T^\mu_{\ \nu, {\rm{EM}}} = b^2 u^\mu u_\nu + \dfrac{1}{2}b^2\delta^\mu_{\ \nu} - b^\mu b_\nu.
\end{equation}

The radiative stress-energy tensor is obtained via the M1 closure scheme. We include a radiative viscosity term to better approximate the radiation field in the funnel region as in \citet{Sadowski+2015a}. We include the effects of absorption, emission, and scattering via the electron scattering opacity ($\kappa_{\rm{es}}$), free-free absorption opacity ($\kappa_{\rm{a}}$), thermal synchrotron, and thermal Comptonization \citep{Sadowski+2015b,Sadowski+2017}. For the $M=10\, M_\odot$ models, we also account for the bound-free absorption opacity ($\kappa_{\rm{bf}}$) using the Sutherland Dopita model \citep{Sutherland-Dopita1993} assuming a solar metal abundance for the gas\footnote{The $10M_\odot$ models are quite hot, with temperatures $>10^7$K, and so the precise details of the atomic opacity prescription or the choice of metallicity are unimportant.}. We exclude the bound-free absorption opacity for the $M=10^4\, M_\odot$ simulations, because these models are primarily meant to represent rapidly-growing ``heavy'' BH seeds in the early universe that are assumed to form in metal-free halos devoid even of star formation \citep[e.g.,][]{Bromm&Loeb2003}.

We adapt modified Kerr-Schild coordinates with the inner radius of the simulation domain inside of the BH horizon. The uniformly spaced internal coordinates $(x_1,x_2,x_3)$ are related to the Kerr-Schild spherical polar coordinates polar coordinates $(r,\vartheta,\varphi)$ by 
\begin{align}
    r &= e^{x_1},\\
    \vartheta &= \left[1 + \cot\left(\frac{H_0\pi}{2}\right)\tan\left(H_0\pi\left[-0.5 + \left(Y_1 + \dfrac{(-Y_1 + Y_2)}{(e^{x_1}/2)^{P_0}}\right)(1 - 2x_2) + x_2\right]\right)\right]\dfrac{\pi}{2},\\
    \varphi &= x_3.
\end{align}
The complicated form of the middle expression is designed such that (i) the minimum/maximum coordinate $\vartheta$ is radially dependent, and (ii) more cells are focused towards the midplane $\vartheta=\pi/2$. We choose $H_0=0.6$ to add slightly more resolution in the midplane in order to better resolve the accretion disk. We also choose $Y_1=0.0025$, $Y_2=0.025$, and $P_0=1.2$ such that $Y_2\pi<\vartheta<(1-Y_2)\pi$ near the horizon but $Y_1\pi<\vartheta<(1-Y_1)\pi$ further away. This choice ultimately increases the minimum time step and decreases the computational cost of each simulation.

The radial grid cells are spaced logarithmically, and we choose inner and outer radial bounds $R_{\rm{min}}<r_H$ and $R_{\rm{max}}=10^4\,r_g$. We specify $R_{\rm{min}}$ for each model in Table \ref{tab:modelinfo}. We also use a wedge of $\pi/2$ in azimuth instead of the full $2\pi$ in order to minimize computational costs and set $\varphi_{\rm{min}}=-\pi/4$ and $\varphi_{\rm{max}}=\pi/4$. We choose outflow boundary conditions at both the inner and outer radial bounds, reflective boundary conditions at the top and bottom polar boundaries, and periodic boundary conditions in $\varphi$. In each simulation, we employ a resolution of $N_r\times N_\vartheta\times N_\varphi=256\times192\times24$.
The resolution in $\theta$ is especially important for GRRMHD (and also GRMHD) simulations. The $\theta$ resolution used in the present work is superior to most GRRMHD simulations in the literature. Our $\varphi$ resolution is modest: 24 cells over a $\pi/2$ wedge, which corresponds to an effective resolution of 96 cells over $2\pi$. This is a bit lower than 32 cells in the wedge, or 128 cells over $2\pi$, used in  \citet{Narayan+2017}. However, it is superior to most other GRRMHD simulations reported in the literature, e.g., 64 cells over $2\pi$ in \citet{McKinney+2015} and \citet{Takahashi+2016}, or even 32 cells over $2\pi$ used in other work.

We ensure that the fastest growing mode of the magnetorotational instability (MRI, \citealt{Balbus1991}) is adequately resolved within each simulation. For this we compute the quantities \citep{Hawley2011},
\begin{align}
  Q_\vartheta = \dfrac{2\pi}{\Omega\, dx^\vartheta}\dfrac{|b^\vartheta|}{\sqrt{4\pi\rho}}, \label{eq:Qtheta} \\
  Q_\varphi = \dfrac{2\pi}{\Omega\, dx^\varphi}\dfrac{|b^\varphi|}{\sqrt{4\pi\rho}}, \label{eq:Qphi}
\end{align}
where $dx^i$ (the grid cell size) and $b^i$ (the magnetic field strength) are both evaluated in the orthonormal frame, $\Omega$ is the angular velocity, and $\rho$ is the gas density. $Q\geq 5$ is sufficient to resolve the MRI. We weight $Q$ by $\sqrt{b^2\rho}$ and integrate over the disk ($\sigma < 1$). We then spatially average over $r=10r_g-100r_g$ and temporally average over $t=25000t_g-30000t_g$. In our least resolved model, which has $M=10M_\odot, \, a_*=0.97$, and $f_{\rm{Edd}}=1.97$ in \autoref{tab:modelinfo}, we find $\langle Q_\vartheta \rangle=5$ and $\langle Q_\varphi \rangle=47$, which is sufficient to resolve MRI in the bulk of the disk. $Q_\vartheta$ and $Q_\varphi$ increase with $f_{\rm{Edd}}$ since the disk becomes thicker; therefore, all of our models sufficiently resolve the MRI.

We initialize each simulation with a torus of gas in hydrodynamic equilibrium following \cite{Penna+2013a}. The density was fixed by the entropy constant $\mathcal{K}=63$ and assuming $\Gamma=4/3$. The angular velocity
at the equatorial plane was set to a constant fraction of $\xi=0.975$ of
the Keplerian angular velocity outside radius $R_1=30\,r_g$, and followed
fixed angular momentum between $R_{\rm{in}} < r < R_1$ with $R_{\rm{in}}=22\,r_g$ being the inner edge of the torus. The angular momentum was kept constant along the von-Zeipel cylinders. We set the outer edge of the torus at $r\approx 400 \, r_g$. This method only gives the hydrodynamic quantities. To initialize the radiation, we split the total pressure given by the initial hydrodynamics solution into gas and radiation components by assuming local thermodynamic equilibrium (LTE). We assign the gas and radiation pressure by finding the LTE temperature given by
\begin{equation}
    p_{\rm{tot}}=p_{\rm{gas}}+p_{\rm{rad}}=k_{\rm{B}}\rho T + \dfrac{1}{3}a c T^4,
\end{equation}
where $p_{\rm{tot}}$ is the sum of gas and radiation pressure given by the initial torus in pure hydrodynamics, $p_{\rm{gas}}$ is gas pressure, and $p_{\rm{rad}}$ is the radiation pressure.

We thread the torus with a large scale poloidal magnetic field defined by the vector potential $A_\phi$. We adopt a definition of $A_\phi$ which is a function of $r$ and $\vartheta$ given by
\begin{equation}
    A_\phi=q(r,\vartheta)\sin\left(F(r)-F(R_{\rm{start}})\right),
\end{equation}
where we define
\begin{equation}
q(r,\vartheta) = 
     \begin{cases}
       \dfrac{\left(u_g(r,\vartheta)-u_g(R_{\rm{chop}},\pi/2)\right) - 0.2\left(u_g(r,\pi/2)-u_g(R_{\rm{chop}},\pi/2)\right)}{0.8\left(u_g(r,\pi/2)-u_g(R_{\rm{chop}},\pi/2)\right)}\sin(\vartheta)^3, \quad & R_{\rm{start}} < r < R_{\rm{chop}}  \\
       0 , \quad  & r > R_{\rm{chop}} \\
     \end{cases}
\end{equation}
and
\begin{equation}
    F(r)=\dfrac{1}{\lambda}\left(\dfrac{5}{3}r^{0.6}  + \dfrac{5}{4}r^{-0.4} \right).
\end{equation}
We set each of the parameters $R_{\rm{start}}=1.25 R_{\rm{in}}$, $R_{\rm{chop}}=350\,r_g$, and $\lambda=15$. Note that $q(r,\vartheta)$ uses the midplane gas internal energy to scale the vector potential. Also note that the $\sin(F(r)-F(R_{\rm{start}}))$ term can vary the sign of $A_\phi$ across radius with a wavelength that varies with $\lambda$. Our parameter choices are designed to place a large poloidal field that does not vary in sign at all. We normalize the magnetic field strength by setting the pressure ratio $\beta_{\rm{max}}\equiv(2(p_{\rm{gas}}+p_{\rm{rad}})/b^2)_{\rm{max}}=20$. From these initial conditions, the MAD state naturally develops as the magnetic field is advected towards the horizon in the accretion flow.

We artificially increase the gas density in high magnetization, $\sigma\equiv b^2/\rho$, regions in order to ensure the simulation remains numerically stable by limiting $\sigma\leq60$.  Each simulation is carefully inspected to ensure that its accretion rate, magnetic flux parameter, and radial inflow profiles are in steady state for the window considered for further analysis.  See \autoref{tab:modelinfo} for the full list of simulations described in this work.

\begin{table*}
    \centering
    \begin{tabularx}{\textwidth}{@{}l *6{>{\centering\arraybackslash}X}@{}}
        \hline
        \hline 
        $M$ & $a_*$ & $R_{\rm{min}}$ & $f_{\rm{Edd}}$ & $\langle\phi\rangle$ & $\eta_\mathrm{MHD}$ & $s$ \\
                $(M_\odot)$ & & $(r_g)$ & & & & \\
    \hline
    10 & $-0.9$ & 1.25 & 37.1 & 31.1 & .403 & 7.07 \\
     & & & 9.40 & 20.6 & .153 & 6.14 \\
     & & & 3.86 & 9.07 & .057 & 5.67 \\
     & & & & & & \\
    & $-0.68$ & 1.5 & 33.2 & 33.7 & .216 & 5.94 \\
     & & & 7.82 & 24.5 & .122 & 5.60 \\
     & & & 1.81 & 8.76 & .053 & 5.16 \\
     & & & & & & \\
    & $0$ & 1.75 & 10.4 & 56.6 & .148 & 1.20 \\
     & & & 8.09 & 41.1 & .119 & 2.13 \\
     & & & 3.63 & 41.0 & .113 & 2.14 \\
     & & & 1.66 & 23.5 & .078 & 2.90 \\
     & & & 1.40 & 14.3 & .071 & 3.05 \\
     & & & 1.07 & 10.9 & .071 & 3.10 \\
     & & & & & & \\
    & $0.68$ & 1.5 & 18.6 & 54.8 & .564 & -5.42 \\
     & & & 8.26 & 51.2 & .576 & -4.74 \\
     & & & 3.63 & 50.6 & .489 & -4.17 \\
     & & & 2.90 & 28.4 & .213 & -.328 \\
     & & & 1.21 & 17.5 & .137 & .665 \\
     & & & & \\
    & $0.9$ & 1.25 & 24.9 & 53.4 & 1.35 & -8.81 \\
     & & & 11.3 & 50.4 & 1.10 & -7.03 \\
     & & & 8.65 & 38.0 & .615 & -3.48 \\
     & & & 2.60 & 35.9 & .507 & -2.64 \\
     & & & 2.20 & 25.9 & .288 & -.810 \\
     & & & 1.84 & 17.6 & .196 & -.159 \\
     & & & & & & \\
    & $0.97$ & 1.09 & 40.7 & 49.3 & 1.61 & -8.50 \\
     & & & 19.5 & 49.8 & 1.68 & -8.82 \\
     & & & 8.49 & 44.2 & 1.21 & -6.14 \\
     & & & 3.59 & 37.5 & .718 & -3.54 \\
     & & & 1.97 & 18.1 & .244 & -.406 \\
     & & & & & & \\
    $10^4$ & $0$ & 1.75 & 7.94 & 61.1 & .152 & 1.27 \\
     & & & 6.40 & 48.9 & .135 & 1.73 \\
     & & & 3.23 & 43.6 & .118 & 2.05 \\
     & & & 1.37 & 31.2 & .088 & 2.63 \\
     & & & .402 & 22.3 & .074 & 2.88 \\
     & & & & & & \\
    & $0.9$ & 1.25 & 18.2 & 64.6 & 2.69 & -17.0 \\
     & & & 9.27 & 54.3 & 1.38 & -10.1 \\
     & & & 5.34 & 45.7 & .845 & -5.56 \\
     & & & 2.47 & 40.3 & .657 & -3.97 \\
     & & & .925 & 30.3 & .372 & -1.74 \\
    \hline
    \end{tabularx}
    \caption{Description of simulations presented in this work.  Note that $f_\mathrm{Edd}$ and $\langle \phi \rangle$ are computed at the horizon, but $\eta_\mathrm{MHD}$ and $s$ are computed at a radius of $5 \ r_g$.}
    \label{tab:modelinfo}
\end{table*}

\section{Flux Calculations}
\label{sec:fluxes}

The mass accretion rate as a function of radius is computed as
\begin{equation} \label{eq:mdotin}
  \dot{M}(r) = -\int_\vartheta \int_\varphi \sqrt{-g}\rho \,u^r d\varphi d\vartheta.
\end{equation}

As we discuss in \autoref{sec:spin}, we model the hydrodynamic and electromagnetic parts of the spinup parameter separately, following the formalism of \citet{Moderski&Sikora1996} and \citet{Lowell+2023}. To that end, we compute the angular momentum flux normalized by the mass accretion rate in HD and EM components separately:
\begin{equation}   \label{eq:lHD} 
    l_{\rm{HD}}(r) = -\frac{1}{\dot{M}(r)}\int_\vartheta \int_\varphi T^r_{\ \varphi, {\rm{HD}}}\sqrt{-g}\, d\varphi d\vartheta,
\end{equation}
\begin{equation}    \label{eq:lEM}
    l_{\rm{EM}}(r) = -\frac{1}{\dot{M}(r)}\int_\vartheta \int_\varphi T^r_{\ \varphi, {\rm{EM}}}\sqrt{-g}\, d\varphi d\vartheta.
\end{equation}
We similarly obtain the energy flux normalized by the mass accretion rate in HD and EM components:
\begin{equation}    \label{eq:eHD}
    e_{\rm{HD}}(r) = -\frac{1}{\dot{M}(r)}\int_\vartheta \int_\varphi T^r_{\ t, {\rm{HD}}}\sqrt{-g} \,d\varphi d\vartheta,
\end{equation}
\begin{equation}    \label{eq:eEM}
    e_{\rm{EM}}(r) = -\frac{1}{\dot{M}(r)}\int_\vartheta \int_\varphi T^r_{\ t, {\rm{EM}}}\sqrt{-g}\, d\varphi d\vartheta.
\end{equation}
Note that the choice of sign in each expression is such that we compute the flux of energy and angular momentum \textit{into} the BH, both of which are positive. We are particularly interested in the total outflowing energy relative to the accreted rest mass energy. We characterize this numerically using the dimensionless MHD efficiency
\begin{equation}    \label{eq:etaMHD}
    \eta_{\rm{MHD}}(r) = 1 -[e_{\rm{HD}}(r)+e_{\rm{EM}}(r)].
\end{equation}

\begin{figure*}
  \centering
  \includegraphics[width=\textwidth]{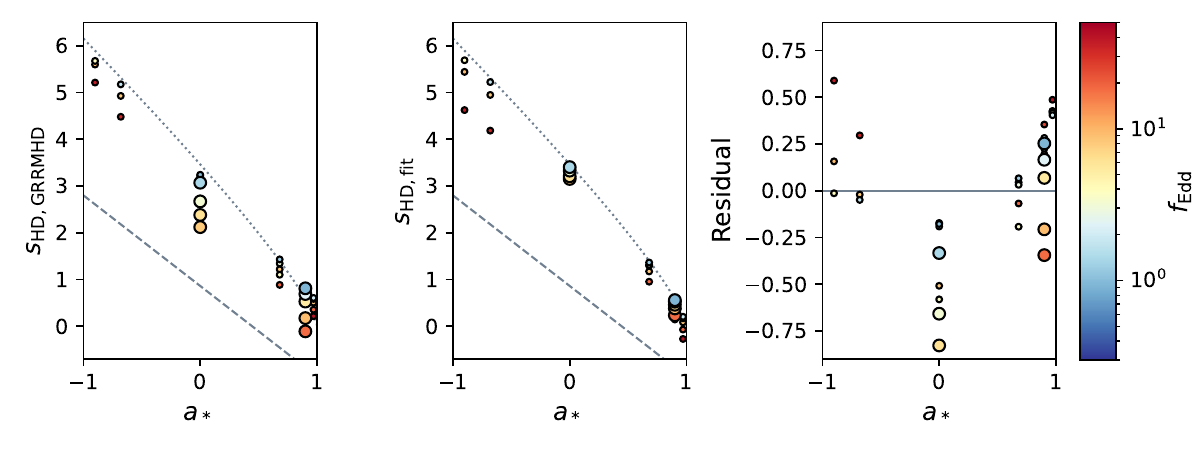}
  \caption{Hydrodynamic spinup parameter calculated from our simulations (left), our fitting function (center), and residuals (right).  The dotted line is the thin disk solution, and the dashed line is the value found by \citet{Lowell+2023} for non-radiative simulations.}
  \label{fig:fluxes_fit}
\end{figure*}

For the hydrodynamic spinup component, we first obtain the specific angular momentum fluxes $l_\mathrm{HD}$ (\autoref{eq:lHD}) and specific energy fluxes $e_\mathrm{HD}$ (\autoref{eq:eHD}) from the fluid simulations at a radius of $5 \ r_g$. We plot the values calculated directly from the GRRMHD simulations in the leftmost panel of \autoref{fig:fluxes_fit}.  The dotted line represents the analytic solution for a thin disk, which we refer to $s_\mathrm{thin}$.  As expected, the models approach $s_\mathrm{thin}$ as $f_\mathrm{Edd} \to 0$.  Meanwhile, the dashed line represents the fit found for non-radiative GRMHD simulations from \citet{Lowell+2023}, which we refer to as $s_\mathrm{min}$.  They reported $e_\mathrm{HD} \approx 0.86$ and $l_\mathrm{HD} \approx 0.97$ independent of spin, and thus 

\begin{equation}
    s_\mathrm{min} = 0.86 - 1.94a_*.  
    \label{eqn:s_min}
\end{equation}

As $f_\mathrm{Edd}$ increases, our simulations appear to move from $s_\mathrm{thin}$ towards $s_\mathrm{min}$.  To build our model, we devise a fitting function that approaches $s_\mathrm{thin}$ as $f_\mathrm{Edd} \to 0$, and $s_\mathrm{min}$ as $f_\mathrm{Edd} \to \infty$.  Thus, we fit for a single number to interpolate between these solutions, arriving at

\begin{equation}
    s_\mathrm{HD} = \frac{s_\mathrm{thin} + s_\mathrm{min} \xi}{1+\xi}
    \label{eqn:sHD}
\end{equation}

\noindent with $\xi = 0.017 \; f_\mathrm{Edd}$.

The results of this fitting function are shown in the central column of \autoref{fig:fluxes_fit}, and residuals are shown in the rightmost column.  Without modeling an additional spin dependence, this fitting function underestimates the rapidity with which the $a_*=0$ models transition from $s_\mathrm{thin}$ to $s_\mathrm{min}$.  We speculate that this may be due to the lack of consideration of angular momentum loss due to a hydrodynamic wind, evident in \autoref{fig:phimdot}.

For convenience, we reproduce the formulae to obtain $s_\mathrm{thin}$ here, following \citet{Moderski&Sikora1996}.  In units where $G=c=M=1$,

\begin{equation}
    e_\mathrm{thin} = \left( 1 - \frac{2}{3 r_\mathrm{ms}} \right)^{1/2}, 
\end{equation}

\noindent and 

\begin{equation}
    l_\mathrm{thin} = \frac{2}{3\sqrt{3}} \left[ 1 + 2(3 r_\mathrm{ms}-2)^{1/2} \right],
\end{equation}

\noindent where $r_\mathrm{ms}$ is the radius of the marginally stable orbit, given by 

\begin{equation}
    r_\mathrm{ms} = 3 + Z_2 - \mathrm{sign}(a_*)[(3-Z_1)(3+Z_1+2Z_2)]^{1/2}, 
\end{equation}

\noindent for 

\begin{equation}
    Z_1 = 1 + (1-a_*^2)^{1/3}[(1+a_*)^{1/3}+(1-a_*)^{1/3}] 
\end{equation}

\noindent and 

\begin{equation}
    Z_2 = (3a_*^2+Z_1^2)^{1/2}.
\end{equation}

\noindent Finally,

\begin{equation}
    s_\mathrm{thin} = l_\mathrm{thin} - 2 a_* e_\mathrm{thin}.
    \label{eqn:s_thin}
\end{equation}

We also use the radiative efficiency of the thin disk model to define the Eddington ratio.  The Eddington luminosity is the limiting luminosity above which radiation pressure exceeds gravitational pressure in a spherically symmetric system.  It is given by

\begin{equation}
   L_\mathrm{Edd} = \frac{4 \pi G M m_p c}{\sigma_T},
\end{equation}

\noindent where $m_p$ is the proton mass and $\sigma_T$ is the Thomson cross-section.  Defining a radiative efficiency $\epsilon = L / \dot{M} c^2$ allows one to define the Eddington mass accretion rate,

\begin{equation}
   \dot{M}_\mathrm{Edd} = \frac{4 \pi G M m_p}{\epsilon \sigma_T c}, 
   \label{eqn:mdot_edd}
\end{equation}

Throughout this work, when defining the Eddington mass accretion rate, we assume the radiative efficiency of a thin disk, given by 

\begin{equation}
    \epsilon = 1 - e_\mathrm{thin} = 1 - \left( 1 - \frac{2}{3 r_\mathrm{ms}} \right)^{1/2}.
    \label{eqn:radiative_efficiency_thin}
\end{equation}

\noindent Thus, our definition of $\dot{M}_\mathrm{Edd}$ depends on both mass and spin.

\section{Pressure Scale Height}
\label{sec:scale_height}

\begin{figure}
  \centering
  \includegraphics[width=\textwidth]{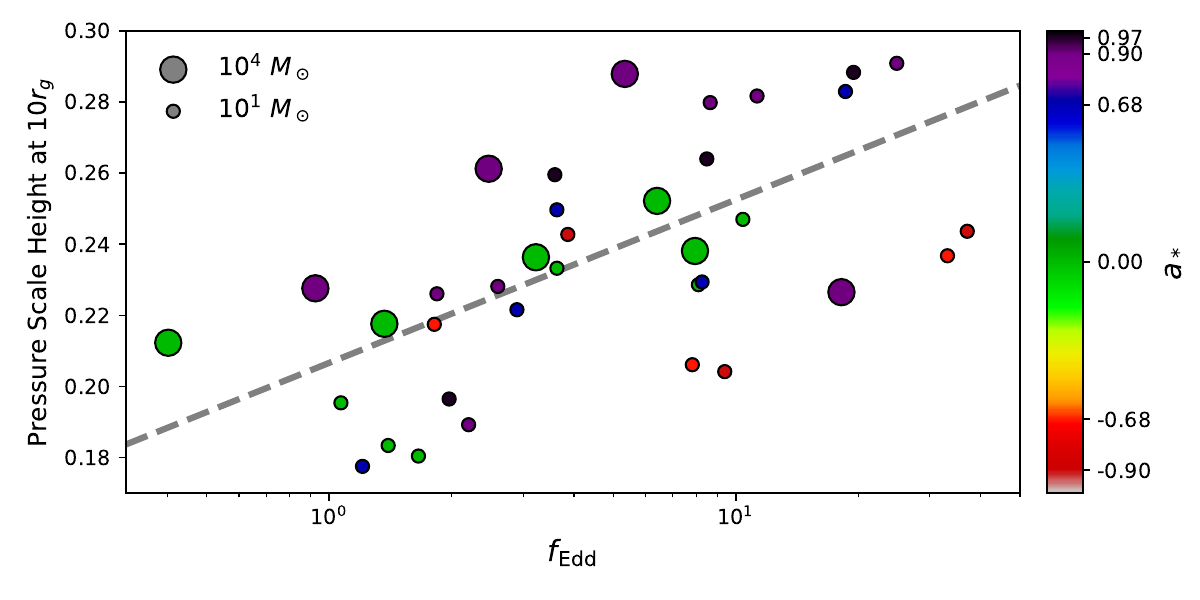}
  \caption{Pressure scale height measured at a radius of $10 \; r_g$ for our simulations, colored by spin.  In grey, we plot a linear regression to these data, $h/r = 0.21 + 0.046\log_{10}f_\mathrm{Edd}$.  This correlation suggests that higher Eddington ratios lead to greater pressures that can better confine magnetic flux on event horizon scales.}
  \label{fig:scale_height}
\end{figure}

To gain greater insight into the link between magnetic flux and Eddington ratio presented in \autoref{fig:phimdot}, we explore the pressure scale heights of our simulations.  We define the pressure scale height to be

\begin{equation}
    \frac{h}{r} = \frac{\int\int (P_\mathrm{gas} + P_\mathrm{rad})|\pi/2-\theta| \sqrt{-g}\,\mathrm{d}\theta\,\mathrm{d}\phi}{\int\int (P_{gas}+P_{rad}) \sqrt{-g}\,\mathrm{d}\theta\,\mathrm{d}\phi},
    \label{eq:hr}
\end{equation}

\noindent where $P_\mathrm{gas}$ is the gas pressure and $P_\mathrm{rad}$ is the radiation pressure (which dominates).  Here, $P_\mathrm{gas} + P_\mathrm{rad}$ has taken the place of $\rho$ in the usual definition of the scale height.  In \autoref{fig:scale_height}, we plot the pressure scale height at a radius of $10 \ r_g$ as a function of Eddington ratio in our simulations, and color code by spin.  In grey, we plot a linear regression to these data, from which we obtain $h/r = 0.21 + 0.046\log_{10}f_\mathrm{Edd}$.  This increase in pressure scale height as a function of Eddington ratio suggests that a higher Eddington ratio results in more pressure, mostly due to radiation, that can drive the gas to confine stronger magnetic fields onto the horizon.

\end{document}